\documentclass[twoside]{articlek}
\usepackage{slashed}
\usepackage{amssymb, amsmath}
\usepackage{physics}

\usepackage{multirow}
\usepackage{dcolumn}% Align table columns on decimal point

\usepackage[normalem]{ulem} %MARTINOVE SKRTANICE
\usepackage{xcolor}%%% MARTINOVE FARBICKY
\definecolor{MGreen}{RGB}{34,139,34}
\renewcommand{\refname}{REFERENCES}
\def\be{\begin{equation}}
\def\ee{\end{equation}}
\def\BibTeX{{\rm B\kern-.05em{\sc i\kern-.025em b}\kern-.08em
            T\kern-.1667em\lower.7ex\hbox{E}\kern-.125emX}}

\usepackage{graphicx}
\begin{document}
\sloppy

%\vspace*{1.7cm}   % for the title  page only
%\begin{center}
\twocolumn[
{\large\bf Numerical modeling of SNSPD absorption utilizing optical conductivity with quantum corrections}\\

{\noindent\small M.~Baránek$^{1}$,  P.~Neilinger$^{1,2}$, S.~Kern$^{1}$, and M.~Grajcar$^{1,2}$\\
%pavol.neilinger@fmph.uniba.sk, martin.baranek@fmph.uniba.sk,
\noindent$^{1}$ Department of Experimental Physics, Comenius University, SK-842 48, Bratislava, Slovakia\\
$^{2}$ Institute of Physics, Slovak Academy of Sciences, D\'{u}bravsk\'{a} cesta 9, SK-845 11, Bratislava, Slovakia
}\\

%\end{center}
%\vspace*{1ex}

{\bf ABSTRACT.} Superconducting nanowire single-photon detectors are widely used in various fields of physics and technology, due to their high efficiency and timing precision. Although, in principle, their detection mechanism offers broadband operation, their wavelength range has to be optimized by the optical cavity parameters for a specific task. We present a study of the optical absorption of a superconducting nanowire single photon detector (SNSPD) with an optical cavity. The optical properties of the niobium nitride films, measured by spectroscopic ellipsometry, were modelled using the Drude-Lorentz model with quantum corrections. The numerical simulations of the optical response of the detectors show that the wavelength range of the detector is not solely determined by its geometry, but the optical conductivity of the disordered thin metallic films contributes considerably. This contribution can be conveniently expressed by the ratio of imaginary and real parts of the optical conductivity. This knowledge can be utilized in detector design.
\vspace{5mm}]

%---------------------------------------------------------------------------

\section{INTRODUCTION}

Strengthening the security of digital communication channels by building Quantum Key Distribution (QKD) infrastructure is an ongoing effort worldwide\cite{Chen2021,Ribezzo23}. Most QKD systems are based on entangled photon pairs as quantum bits and utilize the existing, low-loss fiber optical infrastructure \cite{Aktas}. To implement an efficient algorithm, a single-photon detector with a high overall detection efficiency, low dark counts, and high speed is required\cite{Shibata14}. The State-of-the-art detectors that fulfill these requirements are the superconducting nanowire single-photon detectors (SNSPD)\cite{Goltsman01}. They are superior in terms of detection efficiency and dark count rates to other types\cite{Zadeh21}, such as the Avalanche Photodiodes (APD)\cite{Renker06}, and do not require sub-kelvin temperatures, as the Transition Edge Sensors (TES)\cite{Cabrera98} do. Moreover, their superior properties make them beneficial in a broad range of applications in different areas of physics, for example, physical chemistry and spectroscopy\cite{Lau23}, fluorescent luminescence\cite{Wang19}, or fast space-to-ground communication\cite{Hao24} often requiring a lower wavelength range. The detector's wavelength can be optimized by its design depending on the required spectral range.  This is also important in the implementation of modern quantum networks\cite{Aktas,Lim10} as not just the detection efficiency of the detector, but its bandwidth is a relevant parameter.

\begin{figure} [h]                     %instead of \begin{figure}[t]
\begin{center}                        %instead of \begin{center}
\includegraphics[width=30mm]{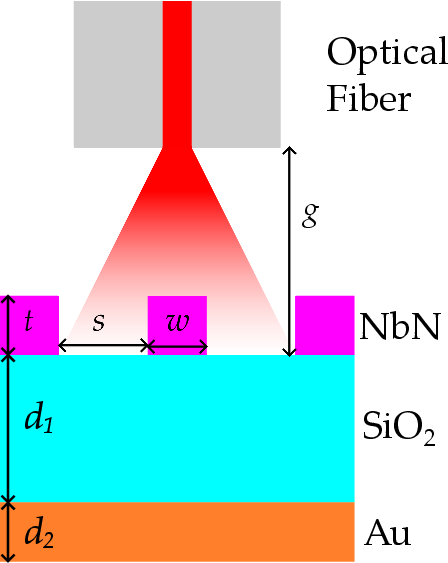}%
\includegraphics[width=30mm]{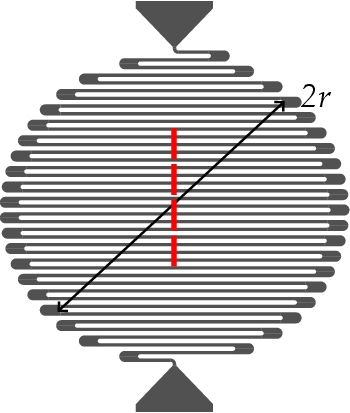}
\end{center}                         %instead of \end{center}
\vspace{-2mm} \caption{Scheme of the SNSPD with coupled optical fiber. Nanowire has thickness $t$, width $w$, and spacing $s$. Optical $\lambda$/4 resonator consists of a dielectric spacer with thickness $d_{1}$ and a gold reflector with $d_{2}$. The gap between optical fiber and nanowire is $g$. Right: Top-down view of the nanowire of diameter $2r$, with red line showing the position of the cross-sectional view.}
\label{fig:model}
\end{figure}

The working principle of SNSPD is straightforward: a nanowire in the superconducting state is current-biased close to the edge of the normal metal transition. The absorption of a single photon results in the destruction of superconductivity at some place along the nanowire. Due to the current bias, the normal state region expands, which results in a voltage pulse across the nanowire.  
These nanowires are fabricated on 6-15~nm thick, highly disordered superconductors, such as Niobium Nitride\cite{Zhang17}, Titanium Niobium Nitride\cite{Chang21}, Tungsten Silicide\cite{WSi} and their width is in the range of 30-200~nm. 

The overall system detection efficiency (SDE) of detectors is determined by the efficiency of the photon coupling to the active area of the detector $\eta_{\mathrm{coupling}}$, the photon absorption efficiency of the superconductor $\eta_{\mathrm{abs}}$, and by the above-described efficiency of transformation of the absorbed photon to electric signal $\eta_{\mathrm{intrinsic}}$ and is given as\cite{You20}:

\begin{equation}
\mathrm{SDE} = \eta_{\mathrm{coupling}}\cdot\eta_{\mathrm{abs}}\cdot\eta_{\mathrm{intrinsic}},
\end{equation}

\begin{figure} [h]                     %instead of \begin{figure}[t]
\begin{center}                        %instead of \begin{center}
\includegraphics[width=0.8\linewidth]{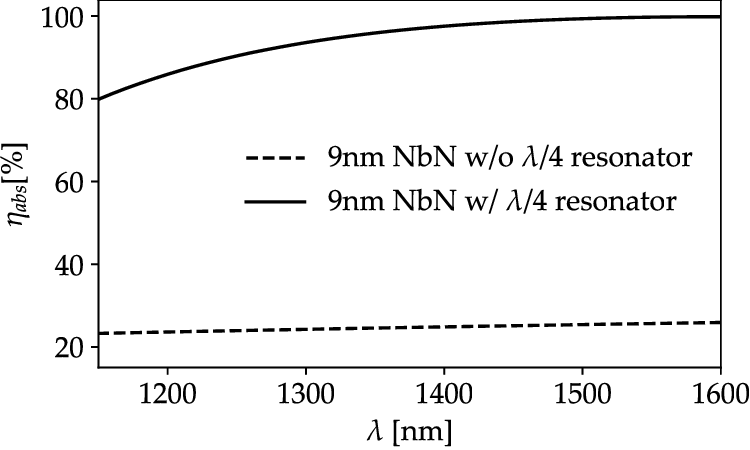}%  &
\end{center}                         %instead of \end{center}
\vspace{-2mm} \caption{Nanowire optical absorption $\eta_{\mathrm{abs}}$ on bare sapphire substrate and on  $\lambda/4$ resonator.}
\label{fig:absorption_basic}
\end{figure}

The absorption of a nanowire (thin film) $\eta_{\mathrm{abs}}$ on bare substrate is insufficient (below 30$\%$, Fig. \ref{fig:absorption_basic}). To increase the absorption at the desired wavelength ($\lambda=$1550~nm) the nanowires are commonly fabricated in an optical resonator, at the cost of limiting its bandwidth. A common choice is a $\lambda/4$ resonant cavity \cite{Redaelli16}. 

This resonator consists of a thin dielectric layer with thickness $d_{1}$, matching its optical length $\Lambda$ with $\lambda/4$ of the desired light's wavelength. The nanowire is fabricated on one side of the dielectric and the other side is covered with a low-loss mirror metallic layer (as shown in Fig. \ref{fig:model}). The resonator enhances the absorption of the photons in the nanowire close to 100$\%$ (Fig. \ref{fig:absorption_basic}) by creating an antinode of the electric field of TE mode in the superconductor. The detectable light is guided by an optical fiber to this structure.

To properly model the optical properties of this structure, and thus to maximize the absorption $\eta_{\mathrm{abs}}$ in the nanowire, not only has the dielectric thickness $d_{1}$ to be optimized, but the precise optical properties of the disordered metal have to be considered. However, their optical properties in the metallic state are not trivial due to the presence of quantum corrections,  which the standard Drude model fails to describe \cite{Neilinger19}. This means that, to model their optical absorption, the optical properties of the specific thin film have to be either measured in the infrared spectra, or the right optical model - or an extrapolation method - has to be chosen, if measurements in the required range are not accessible. 

In this paper, we investigate the dependence of the optical absorption of NbN nanowires in $\lambda/4$ resonant cavity on film thickness and nanowire design. The optical conductivity of a set of NbN films was determined from spectroscopic ellipsometry in the visible range, and they were fitted by the modified Drude-Lorentz model, which takes the quantum corrections into account \cite{Samo_to_be} and provides an excellent fit of these conductivities. The presence of quantum corrections in the IR range results in a strong wavelength and thickness dependence of the optical properties of NbN films. The obtained optical fits and thickness dependencies of its parameters are used to model the absorption spectra of nanowires. The presented approach can be applied to other disordered films, such as MoC and NbTiN. The optical properties of the latter\cite{Banerjee18} are almost identical to the properties of NbN.

\section{Optical properties of NbN films}

Niobium nitride samples were deposited by means of Pulsed Laser Deposition\cite{serhii} from a 99~\% pure niobium target on top of single-side polished sapphire substrate in a nitrogen atmosphere with 1~\% of hydrogen. The film thickness varied from 8 up to 22~nm. The optical conductivities of our films were determined from spectroscopic ellipsometry measured in wavelength range from 400 to 1000~nm at room temperature. 

\begin{figure}
	\centering
	\includegraphics[origin=c,width=\linewidth]{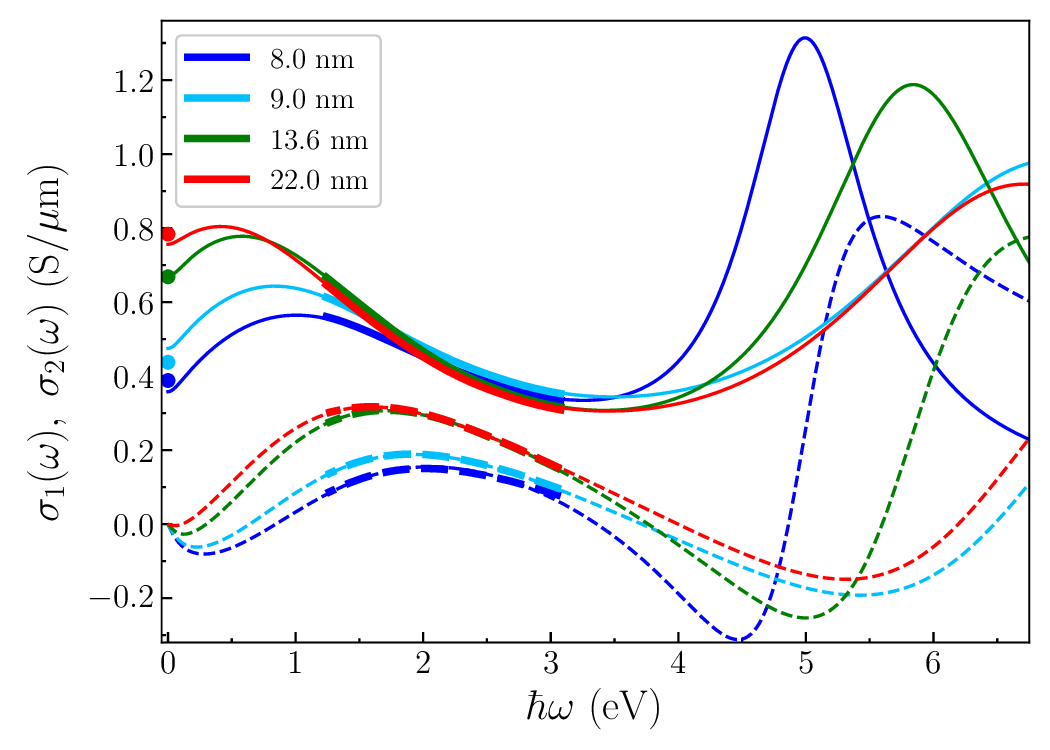}
 	\includegraphics[origin=c,width=\linewidth]{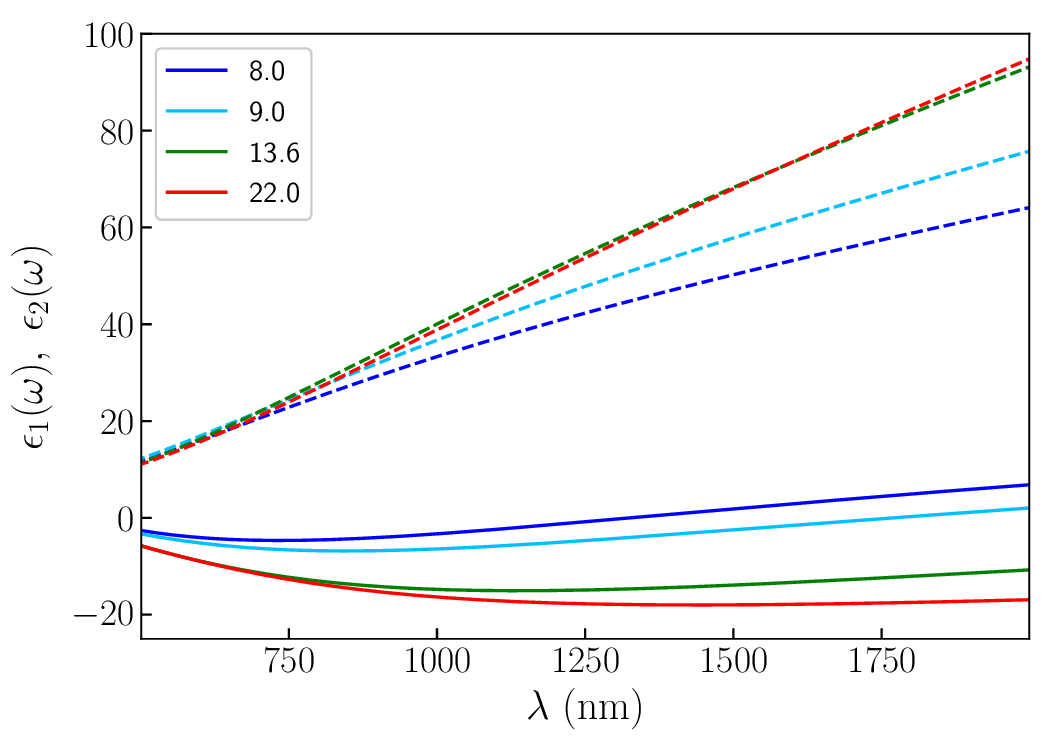}
 
	\caption{Specific conductivity (top) and permittivity (bottom) of thin NbN films, consisting of real (solid line) and imaginary (dashed line) parts. Thick lines are the result of spectroscopic ellipsometry measurement, and thin lines result from the model described above. Dots at zero frequency are the room temperature DC conductivities.}
	\label{fig:Samo_dato}
\end{figure}

\begin{figure} [h]                   
\begin{center}                      
\includegraphics[width=0.8\linewidth]{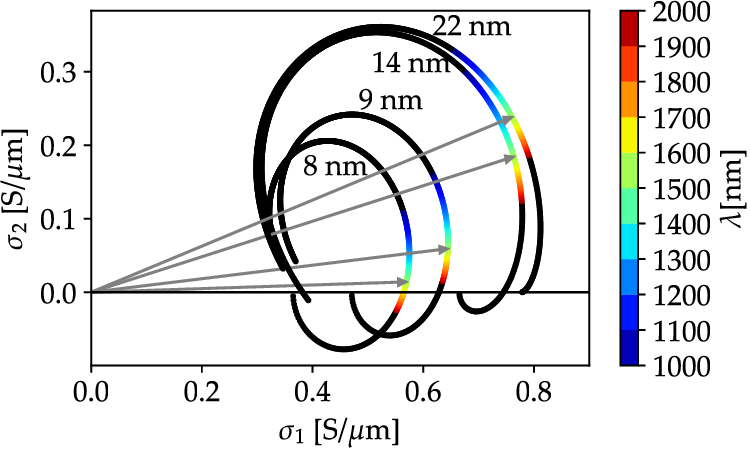}%
\end{center}            
\vspace{-2mm} \caption{Wavelength-dependent real and imaginary part of conductivity for films with various thicknesses. The arrows show the mismatch in the simple rescaling of the conductivities, easily visible by the change in the angle $\theta_0$ of arrows, which point at the conductivity at 1550~nm.}
\label{fig:spiral}
\end{figure}

In Ref.\cite{Samo_to_be} the quantum corrections to the optical conductivity of these NbN films were thoroughly studied. These corrections are arising from localization and interaction effects both having a square root energy dependence\cite{Efros2012electron}. The optical conductivity  $\tilde{\sigma} = \sigma_1 + i\sigma_2$ is described by the modified Drude-Lorentz model in the form:
\begin{equation}
\begin{aligned}
\sigma_1(\omega) =& \frac{\sigma_D}{1+(\frac{\Omega}{\Gamma})^2}\Big(1-\mathcal{Q}^2(1-\sqrt{\frac{\Omega}{\Gamma}})e^{-\frac{1}{2}(\frac{\Omega}{\Gamma})^2}\Big) +\\ &\frac{\sigma_L}{1+\left(\frac{\Omega_L^2-\Omega^2}{\Omega \Gamma_L}\right)^2},
\end{aligned}
\label{eq:TheModel}
\end{equation}

\begin{equation}
\begin{aligned}
\sigma_2(\omega) = \mathcal{H}[\sigma_1(\omega)] - (\varepsilon_\infty-1)\varepsilon_0\omega
\end{aligned}
\label{eq:Imagin}
\end{equation}
respectively.
Here, $\sigma_D$ is the Drude conductivity, $\Gamma$ is the electron relaxation rate and $\mathcal{Q}$ is the strength of the quantum corrections, also referred to as quantumness. The second term in eq. \ref{eq:TheModel} describes the transition peak at $\Omega_L$, with strength $\sigma_L$, and width $\Gamma_L$. This formula describes the anomalous suppression in the IR range and below, which is characteristic of these films.  The optical conductivity for films with different thicknesses is shown in Fig.\ref{fig:Samo_dato} The same result can be represented by $\sigma_1$ vs. $\sigma_2$ plots in Fig.\ref{fig:spiral}, where the wavelength dependence is color-coded. The solid spirals are the fits with eqs. \ref{eq:TheModel},\ref{eq:Imagin}. The fitted parameters are listed in Tab. \ref{tab:params}. 

As can be seen in Fig. \ref{fig:Samo_dato}a, the Drude conductivity $\sigma_0$ varies weakly with thickness, but the relaxation rate $\Gamma$ and the quantumness $\mathcal{Q}$ are strongly thickness dependent. This reflects the commonly observed suppression of DC conductivity and, most importantly, the significant changes in the optical properties of films, especially in the IR range. By lowering the thickness, the real part of the IR range conductivity is suppressed and the imaginary part changes its sign. The dielectric function 
\begin{equation}
    \varepsilon_r(\omega) = \varepsilon_1 + i\varepsilon_2 = 1+i\tilde{\sigma}(\omega)/(\epsilon_0\omega)
\end{equation}
for various film thicknesses is plotted in Fig.\ref{fig:Samo_dato}b. The known thickness dependence of these parameters allows us to conveniently model the optical properties of SNSPDs. The complex refraction index can be expressed in terms of permittivity, as:
\begin{equation}
    \tilde{n} = \sqrt{\varepsilon_r(\omega)},
    \label{eq:n}
\end{equation}

\begin{table*}[h!]
	\centering
	\renewcommand{\arraystretch}{1.5}
	\begin{tabular}{>{\centering\arraybackslash}p{0.8cm} |>{\centering\arraybackslash}p{1.5cm} >{\centering\arraybackslash}p{1.0cm} >{\centering\arraybackslash}p{0.7cm} >{\centering\arraybackslash}p{1.5cm} >{\centering\arraybackslash}p{1.1cm}
			>{\centering\arraybackslash}p{1.0cm} }
		\hline
		$d[\textrm{nm}]$& $\sigma_D  [M S/\textrm{m}]$ & $\hbar\Gamma[\textrm{eV}]$ &  $\mathcal{Q}$& $\sigma_1 [M S/\textrm{m}]$ & $\hbar\Gamma_1[\textrm{eV}]$& $\hbar\Omega_1[\textrm{eV}]$\\ 
		\hline\hline
		8  & 0.77 & 2.1 & 0.7 & 0.9 & 1.0 & 6.13\\ 
		9 & 0.84 & 2.0 & 0.6 & 0.95 & 1.1 & 6.14\\
		14 & 1.00 & 1.76 & 0.66 & 1.12 & 0.56 & 5.19\\
		22 & 0.94 & 1.77 & 0.50 & 1.07 & 0.68 & 5.54\\
		\hline
	\end{tabular} %zaokruhlene hrubky
	\caption{Parameters of optical model eq.(\ref{eq:TheModel}) obtained from the ellipsometric data fit.}
	\label{tab:params}
\end{table*}

It is important to note, that not just the magnitude of the conductivity changes with the thickness, but the ratio of real to imaginary parts changes as well. This is easily visible by the change of the angle of the arrow pointing to the conductivity at $\lambda_0=$1550~nm in Fig. \ref{fig:spiral}. This angle is defined by the tangent of real and imaginary parts of the optical conductivity as $\tan{\theta_0}=\sigma_2(\lambda_0)/\sigma_1(\lambda_0)$ and is a characteristic of the films for a given thickness - it does not depend on the fill factor of the nanowire, since scaling the conductivity scales both real and imaginary part. 

\section{Optical absorption simulations}
We studied the effect of the film's conductivity on the absorption spectra of SNSPD with a similar detector design to \cite{Chang21}. The numerical simulations were carried out in a commercial frequency-domain EM solver\cite{COMSOL}. 
The air gap between the fiber and the surface of the detector leads to additional Fabry-Perót resonances superposing the absorption spectra of the $\lambda/4$ resonator.
This could be expressed in the wavelength dependence of the coupling parameter $\eta_{\mathrm{coupling}}(\lambda)$. In our 2D simulations, the optical fiber output is modelled as a plane wave coupled to air and we assume infinite size of the detector in the plane perpendicular to the light propagation, which corresponds to the ideal coupling $\eta_{\mathrm{coupling}}=1$ and allows us to focus on the absorption in the nanowire $\eta_{\mathrm{abs}}$.

\begin{figure} [h]
\begin{center}
\includegraphics[width=0.9\linewidth]{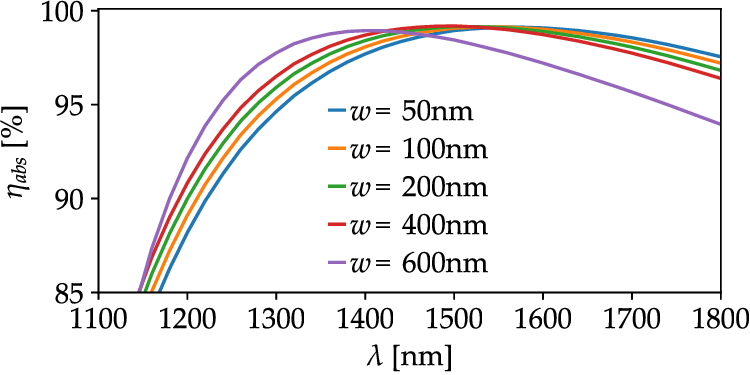}%   
\end{center}                         
\vspace{-2mm} \caption{Simulated absorption spectra of the detector for different nanowire widths and constant fill factor $f = 0.5$ }
\label{fig:skalovanie}
\end{figure}

The optical absorption of the SNSPD can be described by the effective impedance model\cite{Driessen09}, which approximates the thin superconducting meander as a lumped element effective homogeneous medium, with effective sheet conductance $\tilde{G}=G_1+ iG_2$:

\begin{equation}
    \tilde{G} = \tilde{\sigma}\frac{w}{w+s} t \mathrm{\ [S/\square]},
\end{equation}
where $\tilde{\sigma}(t)$ is the optical conductivity (in units of S/m) of the NbN film with thickness $t$, and the ratio ${f = {w}/{(w+s)}}$ is the fill factor.
This approximation is valid for the width and the thickness of the nanowires $w,t \ll \lambda$. As these parameters are usually $w\sim$~30-200~nm, $t\sim$~5-22~nm, and $s\sim$~50-300~nm, this approximation holds well for $\lambda > 1\mu\mathrm{m}$. The validity of this approximation is demonstrated by numerical modeling of the absorption spectra of a detector with different nanowire widths and constant fill factor $f = 0.5$, shown in Fig. \ref{fig:skalovanie}. The absorption spectra are almost identical for $w\lesssim400\mathrm{nm}$. The negligible differences in the absorption maximum ($<0.5\%$) and its wavelength ($<50$~nm) may originate in simulation mesh discretization. The $\approx$ 1\% losses originate in the Au layer with $d_2=~100$~nm.

\begin{figure} [h]
\begin{center}
\includegraphics[width=0.8\linewidth]{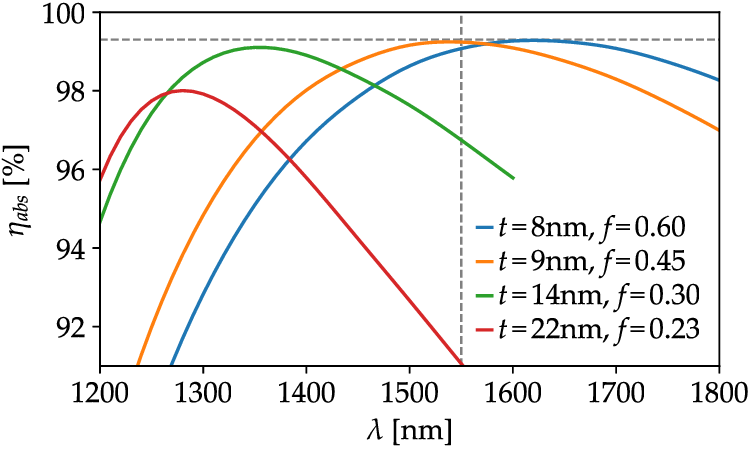}   
\end{center}                         %zo zlozky absorb_seria_true_opted
\vspace{-2mm} \caption{Simulated absorption spectra of nanowires with different thicknesses for fill factor maximizing absorption at 1550~nm.}
\label{fig:absorption_truethick}
\end{figure}

In Fig. \ref{fig:absorption_truethick}, we present the absorption spectra of a detector with 100~nm width NbN nanowire of different thicknesses $t=8-22$~nm on SiO$_2$ layer with thickness $d_1$=~250~nm. The absorption maximum corresponding to the optical length of an ideal $\lambda/4$ resonator is $4\Lambda = 4nd_1 \approx 1600$~nm. 
For each thickness, the fill factor $f$ was tuned to maximize the absorption at $\lambda_0$= 1550~nm. The maximal absorptions at $\lambda_0$, and the wavelength of the absolute maximal absorption $\lambda$ are listed in Table ~\ref{tab:sims_SiO2}. Note, that for this analysis, the optical properties of NbN films were extracted from ellipsometric measurements on films deposited on c-cut sapphire substrate. The optical properties of films deposited on different substrates, for example, SiO$_2$ or Si$_3$N$_4$, can differ. However, the presented results can be easily generalized for any combination of film and dielectric material, as we will show below.
\begin{table}[h!]
	\centering
	\begin{tabular}{|>{\centering\arraybackslash}p{0.8cm} >{\centering\arraybackslash}p{1.5cm} |>{\centering\arraybackslash}p{1.2cm} >{\centering\arraybackslash}p{0.8cm} >{\centering\arraybackslash}p{0.7cm} >{\centering\arraybackslash}p{0.7cm} |}
		\hline
    \multicolumn{6}{|c|}{250~nm SiO$_2$,  $4\Lambda = 1600$~nm}\\
    \hline
		$d[\textrm{nm}]$& $\sigma_2/\sigma_1$  @1550~nm& max($\eta_{abs}$)  $@1550~nm$  [\%] & fill factor & $\lambda_{max}$  [nm] & max  ($\eta_{abs}$)  [\%] \\ 
		\hline
		8  & 0.025 & 99.1 & 0.6 & 1650 & 99.3 \\ 
		9 & 0.092 & 99.2 & 0.45 & 1550 & 99.2 \\
		14 & 0.242 & 96.8 & 0.3 & 1350 & 99.1 \\
		22 & 0.314 & 91.5 & 0.225 & 1250 & 97.9\\
		\hline
	\end{tabular} %zaokruhlene hrubky
	\caption{Simulated maximum absorption at 1550~nm and the optimized fill factor for 1550~nm, wavelength of maximum absorption, and maximum absorption value for different thickness}
	\label{tab:sims_SiO2}
\end{table}

The absorption spectra for $t=$ 8, 9, 14, 22~nm are shown in Fig.~\ref{fig:absorption_truethick}. As it is visible, due to the thickness dependence of the optical conductivity (refraction index), the thickness of the film significantly shifts the position of the absorption maxima and changes the shape of the absorption spectra. The maximum of the absorption value max($\eta_{abs}$), at its peak wavelength $\lambda_{max}$, is also lowered for thicker films due to the increased reflection, but it is negligible compared to the drop of absorption at the desired wavelength $\lambda_{0}$ = 1550~nm. To study how $G_1$ and $G_2$ separately affect the absorption spectra, we simulated the absorption of the nanowire with $w=100$~nm, $s=100$~nm ($f=0.5$), $t=10$~nm on $d_1=250$~nm SiO$_2$ dielectric and $d_2=100$~nm Au reflector. 

\begin{figure}
    \centering
    \includegraphics[width=0.5\linewidth]{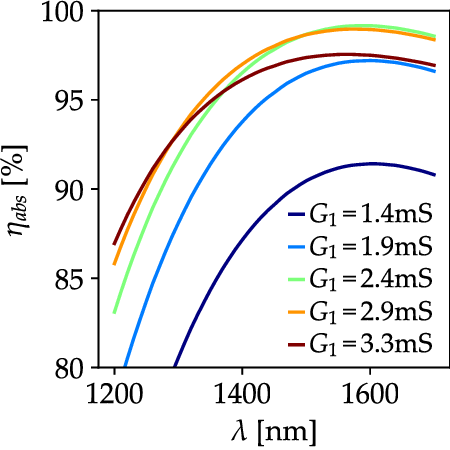}%
    \includegraphics[width=0.5\linewidth]{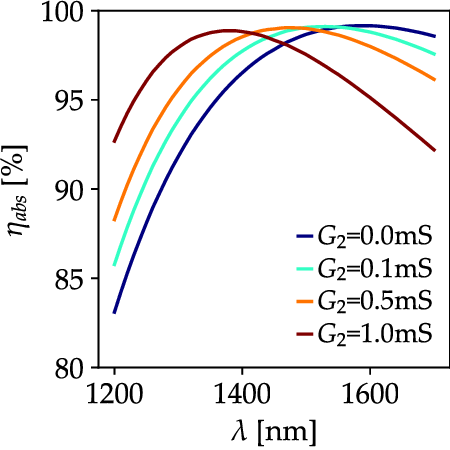}
    
    \caption{Optical absorption in nanowire with fixed $G_2=0\ mS$ (left) and varying $G_1$, and with fixed real effective conductance $G_1 = 2.5\ mS$ (right).}
    \label{fig:absorption_sigma_sweep}
\end{figure}

The absorption spectra for varying $G_1$ and constant $G_2$~=~0 reveal that the real part of the effective sheet conductance $G_1$ ($\sigma_1$, respectively) mainly affects the value of the absorption maxima, but not their position $\lambda_{max}$, shown in Fig.\ref{fig:absorption_sigma_sweep}a. On the other hand, variation of $G_2$($\sigma_2$, respectively) for constant $G_1$ results in the shift of absorption peak $\lambda_{max}$, shown in Fig.\ref{fig:absorption_sigma_sweep}b. The simulated range of $\tan{\theta}$ is approx. 0 - 0.4, which is comparable to the values of our films (0 - 0.32). Thus, the resonant absorption peak of the detector is shifted by the imaginary part of the conductance towards lower values and thus, its wavelength $\lambda_{max}$ is not solely determined by the thickness of the dielectric layer, as one could simply assume. 
This effect, governed by $\tan{\theta}$, is universal and should be present in any thin film with similar values of $\tan{\theta}$.

\begin{figure}[h]
    \centering
    \includegraphics[width=0.5\linewidth]{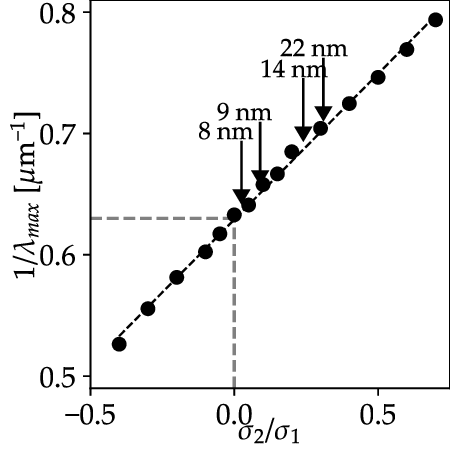}%
    \includegraphics[width=0.5\linewidth]{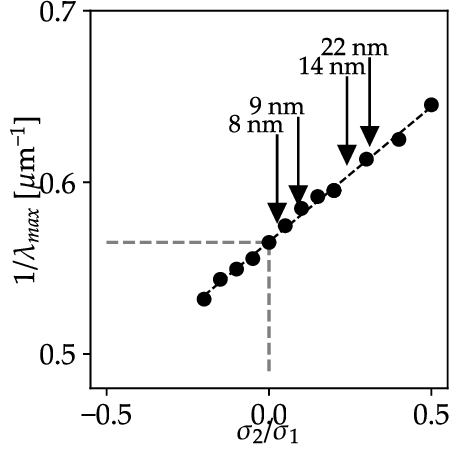}
    
    \caption{Dependance of wavelengths of the simulated absorption maxima on the conductivity ratio $\tan{\theta} = \sigma_2/\sigma_1 = G_2/G_1$, given by samples with different thicknesses. Left: On top of the 250nm SiO$_2$ substrate, Right: On top of the 200~nm Si$_3$N$_4$ membrane}
    \label{fig:absorption_lambda_shift}
\end{figure}

 The dependence of $1/\lambda_{max}$ on the conductivity of the NbN films, represented by $\tan{\theta}$, is linear for a broad range of $\tan{\theta}$, as is visible in Fig.~\ref{fig:absorption_lambda_shift}, and can be approximated as:
\begin{equation}
\frac{1}{\lambda_{max}(\sigma_2)} = {\frac{1}{\lambda_{max}(\sigma_2=0)}+\frac{1}{\lambda_c}\tan\theta},    
\end{equation}
where $\lambda_c$ is a fitting parameter. The value $\lambda_c$ is governed by the optical properties of the dielectric layer and its thickness. For the studied geometry of the detector on SiO$_2$, $\lambda_c\sim4.2$~$\mu$m. Here, we assumed $\tan\theta$ to be constant and equal to $\tan\theta_0$, which is a feasible approximation in the vicinity of $\lambda_0$, as is shown in Fig. \ref{fig:absorption_fixed}.

Since $\tan{\theta}$ increases with decreasing wavelength, the positive feedback further lowers the resonance wavelength.

This simple relation can be directly implemented in the detector design. The absorption maximum shift due to the imaginary part of the conductivity (contribution of the thin film), for a given dielectric layer, limits the possible range of film thicknesses. 

However, using e.g. a commercially available 200~nm thick Si$_3$N$_4$ membrane, the resonance wavelength determined by the membrane is $4\Lambda \approx 1800$~nm and the corresponding absorption at $\lambda_0$ is thus lower, as can be seen in Fig. \ref{fig:absorption_lambda_shift}. Using thicker NbN films, the resonance wavelength can be lowered and the absorption can be significantly increased, see Tab.\ref{tab:sims_SiN2}.  
For each film thickness, the fill factor was optimized to obtain the maximum absorption at $\lambda_0$. Thus, fine-tuning the absorption peak position up to 200~nm is possible, without significant loss of total optical absorption $\eta_{abs}$. 

\begin{table}[h!]
	\centering
	\begin{tabular}{|>{\centering\arraybackslash}p{0.8cm} >{\centering\arraybackslash}p{1.5cm} |>{\centering\arraybackslash}p{1.2cm} >{\centering\arraybackslash}p{0.8cm} >{\centering\arraybackslash}p{0.7cm} >{\centering\arraybackslash}p{0.7cm} |}
		\hline
    \multicolumn{6}{|c|}{200~nm Si$_3$N$_4$,  $4\Lambda = 1800$~nm}\\
    \hline
		$d[\textrm{nm}]$& $\sigma_2/\sigma_1$  @1550~nm& max($\eta_{abs}$)  $@1550~nm$  [\%] & fill factor & $\lambda_{max}$  [nm] & max  ($\eta_{abs}$)  [\%] \\ 
		\hline
		8  & 0.025  & 93.3 & 0.65 & 1800 & 98.8\\ 
		9 & 0.092  & 95.3 & 0.5 & 1750 & 98.7\\
		14 & 0.242 & \textbf{98.6} & 0.3 & 1600 & 98.6\\
		22 & 0.314 & 97.5 & 0.225 & 1450 & 98.1\\
		\hline
	\end{tabular} %zaokruhlene hrubky
	\caption{Table of sample parameters with simulated maximum absorption at 1550~nm, optimized fill factor for 1550~nm, the wavelength of maximum absorption, and maximum absorption value, respectively.}
	\label{tab:sims_SiN2}
\end{table}

The fit parameter for the 200 nm Si$_3$N$_4$ membrane is $\lambda_c = 6.3\mu$m, estimated from the linear fit in Fig. \ref{fig:absorption_lambda_shift}.\vspace{5mm}\\

\subsection{Discussion}
In similar studies, the properties of a single deposited film are often measured (for example by spectroscopic ellipsometry), and the determined specific conductivity is considered to be thickness-independent in the simulations. Consequently, the nanowire's thickness and width are then tuned to maximize the optical absorption in the nanowire at the aimed wavelength $\lambda_0$. For example, in Ref.~\cite{Anant08}, or Ref.~\cite{Yamashita13} the same refraction index was used for samples with thicknesses ranging from 4 to 10~nm, in case of Ref.~\cite{Zhang17} even from 1 to 16~nm. Rescaling the thickness of a specific film to a requested value results in the inaccuracy of the simulated absorption. This fact is acknowledged in \cite{Driessen09}, however, no further analysis is present. Rarely, the refractive index is also considered to be wavelength independent\cite{Anant08}. 

\begin{figure} [h]
\begin{center}
\includegraphics[width=0.8\linewidth]{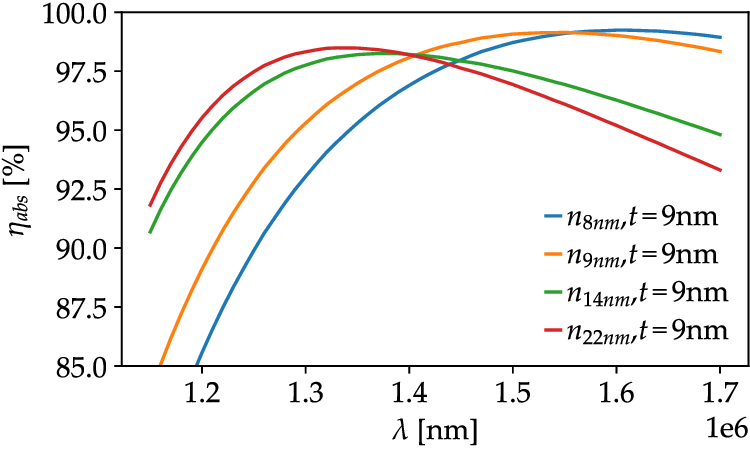}
\end{center}
\vspace{-6mm} \caption{
Simulated optical absorption of 9~nm thick sample on top of the resonant cavity. Other curves show the change in optical absorption of the optical conductivity (refraction index) taken from a sample with different thicknesses.}
\label{fig:absorption}
\end{figure}

To illustrate the error arising from this approximation, we simulated the optical absorption in the above-described nanowire for the respective conductivity $\tilde{\sigma}(t)$, and the corresponding optical indices, given by eq. \ref{eq:n}. For example, using the conductivity of the film with $t_{meas}=$14~nm and simply rescaling it to $t=9$~nm for the simulation would result in a decrease of the simulated absorption maxima by approximately 200~nm, as can be seen in  Fig.\ref{fig:absorption}. 

\begin{figure}
    \centering
    \includegraphics[width=0.8\linewidth]{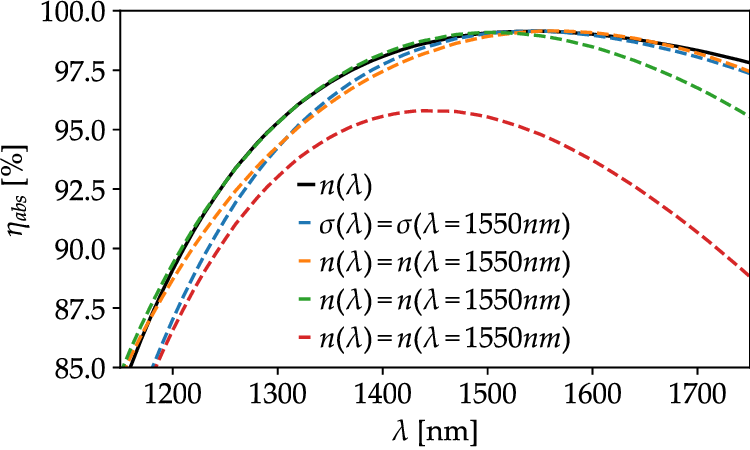}%
    
    \caption{Simulated absorption of a sample with wavelength dependent $n(\lambda)$, constant $n(\lambda=1000,1300,1550\mathrm{nm})$ and $n$ corresponding to constant $\sigma(\lambda=1550\mathrm{nm})$, respectively.}
    \label{fig:absorption_fixed}
\end{figure}

If one assumes the optical conductivity (refraction index) of these films to be wavelength-independent, the error in the absorption spectra will accumulate with the distance from the reference wavelength, as is shown in Fig.~\ref{fig:absorption_fixed}.

\newpage
\section{CONCLUSION}
We modelled the absorption spectra of NbN nanowires in a $\lambda/4$ resonator for different film thicknesses and showed that the optical conductivity of thin films significantly affects the absorption spectra. The optical conductivity of disordered metallic films is strongly affected by the presence of quantum corrections, resulting in nontrivial wavelength and thickness dependence of the optical conductivity. The real part of the conductivity determines the amplitude of absorption of the nanowire in the $\lambda/4$ optical resonator, whereas the imaginary part contributes to the wavelength shift of the maxima. The wavelength shift of the absorption maxima can be expressed in terms of the ratio of imaginary and real parts of the optical conductivity, which is a characteristic of a film with a particular thickness. This knowledge can be utilized in detector design and optimization of its optical properties. Thus, to properly simulate and optimize the absorption of the detectors at a required wavelength, it is necessary to work with the precise optical conductivity of the particular layers. These can be obtained by direct measurements of the samples in the required range, or by utilizing a model that describes the optical conductivity of the disordered samples well, as presented in Ref. \cite{Samo_to_be}.  
\newpage
\section{ACKNOWLEDGMENT}
\noindent This work was supported by the Slovak Research and Development Agency under the contract APVV-20-0425, and by the project skQCI (101091548), founded by the European Union (DIGITAL) and the Recovery and Resilience Plan of the Slovak Republic.

% \newpage
\bibliographystyle{unsrt}
\bibliography{bibliography}

\begin{thebibliography}{10}

\bibitem{Chen2021}
Yu-Ao Chen, Qiang Zhang, Teng-Yun Chen, Wen-Qi Cai, Sheng-Kai Liao, Jun Zhang, Kai Chen, Juan Yin, Ji-Gang Ren, Zhu Chen, Sheng-Long Han, Qing Yu, Ken Liang, Fei Zhou, Xiao Yuan, Mei-Sheng Zhao, Tian-Yin Wang, Xiao Jiang, Liang Zhang, Wei-Yue Liu, Yang Li, Qi~Shen, Yuan Cao, Chao-Yang Lu, Rong Shu, Jian-Yu Wang, Li~Li, Nai-Le Liu, Feihu Xu, Xiang-Bin Wang, Cheng-Zhi Peng, and Jian-Wei Pan.
\newblock An integrated space-to-ground quantum communication network over 4,600 kilometres.
\newblock {\em Nature}, 589(7841):214--219, Jan 2021.

\bibitem{Ribezzo23}
Domenico Ribezzo, Mujtaba Zahidy, Ilaria Vagniluca, Nicola Biagi, Saverio Francesconi, Tommaso Occhipinti, Leif~K. Oxenløwe, Martin Lončarić, Ivan Cvitić, Mario Stipčević, Žiga Pušavec, Rainer Kaltenbaek, Anton Ramšak, Francesco Cesa, Giorgio Giorgetti, Francesco Scazza, Angelo Bassi, Paolo De~Natale, Francesco~Saverio Cataliotti, Massimo Inguscio, Davide Bacco, and Alessandro Zavatta.
\newblock Deploying an inter-european quantum network.
\newblock {\em Advanced Quantum Technologies}, 6(2):2200061, 2023.

\bibitem{Aktas}
Djeylan Aktas, Bruno Fedrici, Florian Kaiser, Tommaso Lunghi, Laurent Labonté, and Sébastien Tanzilli.
\newblock Entanglement distribution over 150 km in wavelength division multiplexed channels for quantum cryptography.
\newblock {\em Laser \& Photonics Reviews}, 10(3):451--457, 2016.

\bibitem{Shibata14}
Hiroyuki Shibata, Toshimori Honjo, and Kaoru Shimizu.
\newblock Quantum key distribution over a 72db channel loss using ultralow dark count superconducting single-photon detectors.
\newblock {\em Opt. Lett.}, 39(17):5078--5081, Sep 2014.

\bibitem{Goltsman01}
G.~N. Gol’tsman, O.~Okunev, G.~Chulkova, A.~Lipatov, A.~Semenov, K.~Smirnov, B.~Voronov, A.~Dzardanov, C.~Williams, and Roman Sobolewski.
\newblock {Picosecond superconducting single-photon optical detector}.
\newblock {\em Applied Physics Letters}, 79(6):705--707, 08 2001.

\bibitem{Zadeh21}
Iman Esmaeil~Zadeh, J.~Chang, Johannes W.~N. Los, Samuel Gyger, Ali~W. Elshaari, Stephan Steinhauer, Sander~N. Dorenbos, and Val Zwiller.
\newblock {Superconducting nanowire single-photon detectors: A perspective on evolution, state-of-the-art, future developments, and applications}.
\newblock {\em Applied Physics Letters}, 118(19):190502, 05 2021.

\bibitem{Renker06}
D.~Renker.
\newblock Geiger-mode avalanche photodiodes, history, properties and problems.
\newblock {\em Nuclear Instruments and Methods in Physics Research Section A: Accelerators, Spectrometers, Detectors and Associated Equipment}, 567(1):48--56, 2006.
\newblock Proceedings of the 4th International Conference on New Developments in Photodetection.

\bibitem{Cabrera98}
B.~Cabrera, R.~M. Clarke, P.~Colling, A.~J. Miller, S.~Nam, and R.~W. Romani.
\newblock {Detection of single infrared, optical, and ultraviolet photons using superconducting transition edge sensors}.
\newblock {\em Applied Physics Letters}, 73(6):735--737, 08 1998.

\bibitem{Lau23}
Jascha~A. Lau, Varun~B. Verma, Dirk Schwarzer, and Alec~M. Wodtke.
\newblock Superconducting single-photon detectors in the mid-infrared for physical chemistry and spectroscopy.
\newblock {\em Chem. Soc. Rev.}, 52:921--941, 2023.

\bibitem{Wang19}
Heqing Wang, Hao Li, Lixing You, Yong Wang, Lu~Zhang, Xiaoyan Yang, Weijun Zhang, Zhen Wang, and Xiaoming Xie.
\newblock Fast and high efficiency superconducting nanowire single-photon detector at 630 nm wavelength.
\newblock {\em Appl. Opt.}, 58(8):1868--1872, Mar 2019.

\bibitem{Hao24}
H.~Hao, QY. Zhao, and YH. et~al. Huang.
\newblock A compact multi-pixel superconducting nanowire single-photon detector array supporting gigabit space-to-ground communications.
\newblock {\em Light: Science {\&} Applications}, 13(1):25, Jan 2024.

\bibitem{Lim10}
Han~Chuen Lim, Akio Yoshizawa, Hidemi Tsuchida, and Kazuro Kikuchi.
\newblock Wavelength-multiplexed entanglement distribution.
\newblock {\em Optical Fiber Technology}, 16(4):225--235, 2010.

\bibitem{Zhang17}
WeiJun Zhang, LiXing You, Hao Li, Jia Huang, ChaoLin Lv, Lu~Zhang, XiaoYu Liu, JunJie Wu, Zhen Wang, and XiaoMing Xie.
\newblock Nbn superconducting nanowire single photon detector with efficiency over 90{\%} at 1550 nm wavelength operational at compact cryocooler temperature.
\newblock {\em Science China Physics, Mechanics {\&} Astronomy}, 60(12):120314, Oct 2017.

\bibitem{Chang21}
J.~Chang, J.~W.~N. Los, J.~O. Tenorio-Pearl, N.~Noordzij, R.~Gourgues, A.~Guardiani, J.~R. Zichi, S.~F. Pereira, H.~P. Urbach, V.~Zwiller, S.~N. Dorenbos, and I.~Esmaeil~Zadeh.
\newblock {Detecting telecom single photons with 99.5-2.07+0.5\% system detection efficiency and high time resolution}.
\newblock {\em APL Photonics}, 6(3):036114, 03 2021.

\bibitem{WSi}
F.~{Marsili}, V.~B. {Verma}, J.~A. {Stern}, S.~{Harrington}, A.~E. {Lita}, T.~{Gerrits}, I.~{Vayshenker}, B.~{Baek}, M.~D. {Shaw}, R.~P. {Mirin}, and S.~W. {Nam}.
\newblock {Detecting single infrared photons with 93\% system efficiency}.
\newblock {\em Nature Photonics}, 7(3):210--214, March 2013.

\bibitem{You20}
Lixing You.
\newblock Superconducting nanowire single-photon detectors for quantum information.
\newblock {\em Nanophotonics}, 9(9):2673--2692, 2020.

\bibitem{Redaelli16}
L~Redaelli, G~Bulgarini, S~Dobrovolskiy, S~N Dorenbos, V~Zwiller, E~Monroy, and J~M Gérard.
\newblock Design of broadband high-efficiency superconducting-nanowire single photon detectors.
\newblock {\em Superconductor Science and Technology}, 29(6):065016, may 2016.

\bibitem{Neilinger19}
P.~Neilinger, J.~Gregu\ifmmode~\check{s}\else \v{s}\fi{}, D.~Manca, B.~Gran\ifmmode \check{c}\else \v{c}\fi{}i\ifmmode~\check{c}\else \v{c}\fi{}, M.~Kop\ifmmode~\check{c}\else \v{c}\fi{}\'{\i}k, P.~Szab\'o, P.~Samuely, R.~Hlubina, and M.~Grajcar.
\newblock Observation of quantum corrections to conductivity up to optical frequencies.
\newblock {\em Phys. Rev. B}, 100:241106, Dec 2019.

\bibitem{Samo_to_be}
Samuel Kern, Pavol Neilinger, Magdal{\'e}na Pol{\'a}{\v{c}}kov{\'a}, Martin Bar{\'a}nek, Tom{\'a}{\v{s}} Plecenik, Tom{\'a}{\v{s}} Roch, and Miroslav Grajcar.
\newblock Optical and transport properties of nbn thin films revisited.
\newblock {\em arXiv preprint arXiv:2405.03704}, 2024.

\bibitem{Banerjee18}
Archan Banerjee, Robert~M. Heath, Dmitry Morozov, Dilini Hemakumara, Umberto Nasti, Iain Thayne, and Robert~H. Hadfield.
\newblock Optical properties of refractory metal based thin films.
\newblock {\em Opt. Mater. Express}, 8(8):2072--2088, Aug 2018.

\bibitem{serhii}
Serhii Volkov, Maros Gregor, Tomas Roch, Leonid Satrapinskyy, Branislav Grančič, Tomas Fiantok, and Andrej Plecenik.
\newblock Superconducting properties of very high quality nbn thin films grown by pulsed laser deposition.
\newblock {\em Journal of Electrical Engineering}, 70(7):89--94, 2019.

\bibitem{Efros2012electron}
Alex~L Efros and Michael Pollak.
\newblock {\em Electron-electron interactions in disordered systems}.
\newblock Elsevier, 2012.

\bibitem{COMSOL}
COMSOL Multiphysics® v. 6.2. www.comsol.com. COMSOL AB, Stockholm, Sweden.

\bibitem{Driessen09}
{Driessen, E. F. C.}, {Braakman, F. R.}, {Reiger, E. M.}, {Dorenbos, S. N.}, {Zwiller, V.}, and {de Dood, M. J. A.}
\newblock Impedance model for the polarization-dependent optical absorption of superconducting single-photon detectors.
\newblock {\em Eur. Phys. J. Appl. Phys.}, 47(1):10701, 2009.

\bibitem{Anant08}
Vikas Anant, Andrew~J. Kerman, Eric~A. Dauler, Joel K.~W. Yang, Kristine~M. Rosfjord, and Karl~K. Berggren.
\newblock Optical properties of superconducting nanowire single-photon detectors.
\newblock {\em Opt. Express}, 16(14):10750--10761, Jul 2008.

\bibitem{Yamashita13}
Taro Yamashita, Shigehito Miki, Hirotaka Terai, and Zhen Wang.
\newblock Low-filling-factor superconducting single photon detector with high system detection efficiency.
\newblock {\em Opt. Express}, 21(22):27177--27184, November 2013.

\end{thebibliography}

\end{document}